%% file: Manuscript_R2.tex
\documentclass[11pt,journal]{IEEEtran}
\usepackage{graphicx}
\graphicspath{{figures/}}
\usepackage{amsmath}
\usepackage{amssymb}
\usepackage{afterpage}
\usepackage{verbatim}
\usepackage{psfrag}
\usepackage{color}
\usepackage{colortbl}
\usepackage[dvipsnames]{xcolor}
\usepackage{cite}
\usepackage{lettrine}

\usepackage{setspace}
\usepackage{epsfig}
\usepackage{epstopdf}
\usepackage{footmisc}
\usepackage{fmtcount}
\usepackage{stackrel}
\usepackage{float}
\usepackage{breqn}
\usepackage{enumerate}
\usepackage{subcaption}

\definecolor{airforceblue}{rgb}{0.36, 0.54, 0.66}

\usepackage[ruled,vlined,linesnumbered]{algorithm2e}
\SetKwInput{KwInput}{Input}
\SetKwInput{KwOutput}{Output}
\usepackage{multicol} %% Should not be used with balance package.
\usepackage{flushend}
\begin{document}
%\doublespace
 \title {Edge Computing in IoT: A 6G Perspective}
%\title {Edge-assisted IoT services in 6G Networks: Architecture, Deployment, and Insights}

\author{Mariam Ishtiaq, Nasir Saeed,~\IEEEmembership{Senior Member,~IEEE}, Muhammad Asif Khan,~\IEEEmembership{Senior Member,~IEEE},}

% \thanks{This work is supported by Office of Sponsored Research (OSR) at King
% Abdullah University of Science and Technology (KAUST). 
% The authors are with the Department of Electrical Engineering, Computer Electrical and Mathematical Sciences \& Engineering (CEMSE) Division, King Abdullah University of Science and Technology (KAUST), Thuwal, Makkah Province, Kingdom of Saudi Arabia, 23955-6900.}
% }

% \markboth{Submitted to IEEE Internet Computing}{Nasir \MakeLowercase{\textit{et al.}}: Edge Computing in IoT: A 6G Perspective}

\maketitle{}

\begin{abstract} %% words: 141
Edge computing is one of the key driving forces to enable Beyond 5G (B5G) and 6G networks. Due to the unprecedented increase in traffic volumes and computation demands of future networks, multi-access (or mobile) edge computing (MEC) is considered as a promising solution to provide cloud-computing capabilities within the radio access network (RAN) closer to the end users. There has been a significant amount of research on MEC and its potential applications; however, very little has been said about the key factors of MEC deployment to meet the diverse demands of future applications. In this article, we present key considerations for edge deployments in B5G/6G networks including edge architecture, server location and capacity, user density, security etc.  We further provide state-of-the-art edge-centric services in future B5G/6G networks. The paper also present experimental evaluation of edge-based services deployment including video transcoding and deep learning inference.  Lastly, we present some interesting insights and open research problems in edge computing for 6G networks.
\end{abstract}


\section{Introduction}
\IEEEPARstart{B}{eyond} 5G (B5G) and 6G networks are the catalyst to the digital future of humanity and the realization of Industry 5.0. As connectivity and digital transformation accelerate, worldwide industry verticals will increasingly rely on 5G and B5G networks to enable novel applications and services categorized as ultra-low latency, mission-critical reliability, and a high degree of mobility. With Moore’s law, the traditional system on a chip (SoC) approach is evolving to system-technology co-optimization (STCO), and consequent system development methodologies are adopted rapidly.
In this context, B5G/6G networks are envisioned to serve a diverse range of applications by utilizing MEC as a built in service with cloud-native concepts like leveraging operations within and across data centers, communicating in a micro-service environment, and simultaneously providing secure services and applications \cite{Rodrigues2021}. % Please spit the long sentence.

% The realization of B5G/6G technology is credited to the innovation of enabling technologies like millimeter wave (mmWave) and Terahertz (THz) communication, ultra massive multiple-input and multiple-output (U-MIMO), beamforming, and small cells. In addition, the cloud-native B5G architecture offers ubiquitous, convenient, and on-demand network access to a shared pool of configurable computing resources (e.g., networks, servers, storage, applications, and services) that can be rapidly provisioned and released with minimal management effort or service provider interaction. The concepts of encapsulation and reusability make 5G a viable option for enabling connectivity for the vertical industries.

B5G and 6G networks are envisioned to fully and seamlessly integrate various verticals such as Internet of things (IoT), aerial networks (aka drones), satellite access, and under-water communication. To keep up with this staggering expectation, future networks (B5G/6G) will heavily rely on state-of-the-art artificial intelligence (AI) and machine learning (ML) technologies for intelligent network operations and management. In addition to the infrastructural transformations, B5G/6G networks are also expected to support computationally intensive services and applications. \par

To support the two major transformations in network infrastructures and network services, edge computing has gained significant attention and is considered as a built-in service in 6G networks. Although, a huge amount of research literature has been dedicated on MEC capabilities such as caching services and computation offloading techniques, there has been very limited insights on MEC deployment. \par

In this article, we first present the edge computing architecture, evolution, and the underlying core concepts. Then we discuss the key considerations of deploying edge computing infrastructure and services. We then discuss some interesting use-cases of edge-centric services in different industry verticals. Lastly, We conclude the article by presenting open research problems in this realm and some key insights.

%% =========================  Section 2 ==================
\section{Edge Computing in 5G and B5G Networks}
\begin{figure}[!h]
\centering
\includegraphics[width=0.95\columnwidth]{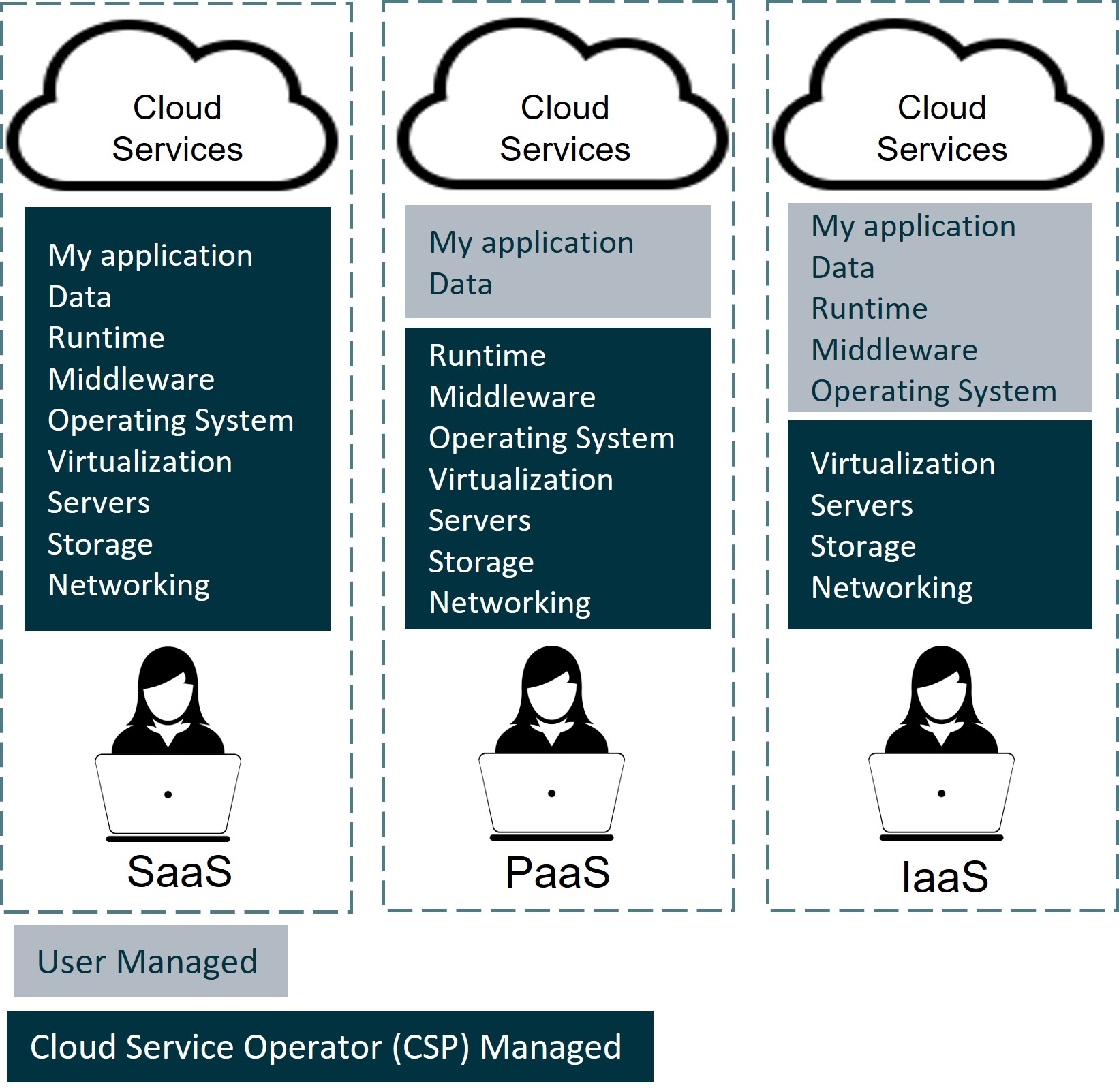}  
\caption{Cloud computing services (CCS).} 
\label{fig:ccs}
\end{figure}

The standard 5G network architecture is built on cloud-native technology enablers like software defined networking (SDN), virtualization, and multi-access edge computing. Conventionally, cloud computing is implemented using the three cloud computing services (CCS) as shown in Fig. \ref{fig:ccs}.

\begin{enumerate}
\item   \textbf{Software-as-a-service (SaaS):} The consumer can use the provider's applications running on a cloud infrastructure where accessibility is possible from various client devices via a thin client interface (e.g., web-based email) or a program interface. In SaaS, the underlying cloud infrastructure, including network, servers, operating systems, and storage, are not managed or controlled by the consumer. 
\item   \textbf{Platform-as-a-service (PaaS): } The consumer can deploy applications created using programming languages, libraries, services, and tools supported by the provider onto the cloud infrastructure. Like SaaS, in PaaS, the consumer does not manage or control the underlying cloud infrastructure, including network, servers, operating systems, or storage. However, in PaaS, the consumer can control the deployed applications and possibly configuration settings for the application-hosting environment.
\item   \textbf{Infrastructure-as-a-service (IaaS):} In IaaS, the consumer can deploy and run arbitrary software, including operating systems and applications, with the provision of processing, storage, networks, and other fundamental computing resources. The underlying cloud infrastructure cannot be managed or controlled by the consumer. However, the consumer can manage the operating systems, storage, deployed applications, and limited networking components (e.g., host firewalls).
\end{enumerate}

One of the shortcomings of traditional cloud computing architecture is that it doesn't meet the ultra-low latency as B5G/6G would commit. The higher delay in cloud computing is mainly caused by the inherited longer distance from the end-users. Edge computing overcomes this by bringing computing and storage capabilities closer to the end-users.

There are different tiers of edge computing architecture: device tier, edge tier, and central tier. The device tier includes devices with limited storage and computing capabilities like sensors, actuators, and user interface devices. This tier often uses proximity networks with limited range and bandwidth. The edge tier comprises intermediate nodes like edge nodes, control nodes, IoT gateways, and other nodes that need low latency or high bandwidth to provide direct support to the nodes in the device tier. The central tier offers a wide connectivity span, potentially using cloud computing services, such as a data center. There are three inherently distributed architectural deployment models for B5G networks. Firstly, cloud computing provides on-demand storage and computing resources. Secondly, edge computing places computing resources closer to the end-users to reduce delay, enhance processing speed, and ensure on-premise security. Also, edge computing gives an added advantage of scalability in terms of increasing users/devices and optimizing reliability and user experience. Finally, MEC is an architectural standard for edge computing by ETSI, providing deployment flexibility at radio nodes, aggregation points, and at the edge of the core network. In addition, reusability (of microservices), extensibility (to include new use cases), and efficiency (in terms of computation time resources and energy) are some of the key features of the MEC standard.

Mobile network operators (MNOs) can provide different deployment strategies. MNOs can move the on-premise edge infrastructure to public or private clouds to ensure enterprise connectivity, including cloud computing, as-a-service connectivity, and MEC infrastructure.  Using such a hybrid approach, communication service providers (CSPs) can pass over cloud platforms (infrastructure and system), focusing entirely on service differentiation. Independent vendors can also give product integration support and customer application support. However, to gain full control of the 5G core, strict service level agreements (SLAs) need to be in place with cloud owners to ensure compliance.

%\begin{figure}[htbp]
%\centering
%\includegraphics[width=0.9\columnwidth]{tex/Edge Computing Architecture.png}  
%\caption{Three-Tiered edge computing architecture is constituted by a device tier, edge tier, and central tier. Three architectural deployment models for 5G networks include cloud computing, edge computing and MEC.} \label{fig:arch}
%\vspace{-1.8em} 
%\end{figure}

%% ==============  Edge Considerations =================
\section{Considerations for Edge-centric B5G/6G Networks}
From MNO's perspective, some of the primary motivations for B5G/6G networks are to deliver high bandwidth and support a high density of users, operate on a cost and energy-efficient infrastructure, and ensure reliability and security of operation. Some of the main considerations for B5G networks to support future services are given in Table \ref{tab:mec2} and discussed in this section.

\begin{enumerate}

\item  \textbf{Edge Architecture and Location:} Edge computing enables networks to become more efficient by providing a virtualization platform, expanding the enterprise service base, and offering network congestion control protocols. From a micro operator's perspective, edge location plays a critical role in ascertaining these core values. In networking terms, edge location refers to its closeness to the consumer. However, this location may vary for operators, enterprises, end-users, and application types, on a need basis. Some of these needs or metrics to decide edge location can be affordable latency, user equipment mobility, network resource management, and computation cost. Various different edge locations can be considered to make the edge availability secure and accessible. Mobile edge servers, retail centers, manufacturing stations, housing compounds, 5G cellular base stations, and the edge of the mobile operator's core network can account for a few edge location possibilities \cite{edgeLocation}. These possibilities can be explored by ensuring an uninterrupted power supply and security to reap the benefits of edge computing.

To meet the stringent QoS requirements of B5G services, MEC promises proximity, a high access rate, and low latency. While there is no definite blueprint for future RAN deployment, distributed unit (DU) and Central unit (CU) functions can be combined or split and deployed at the radio site or in centralized locations. Thus, we can say that location can be determined by services, density, the distance between sites, and transport network speed.

%----------------- Table ----------------------------------
%% For better position in the document
\input{tab_mec2}

\item   \textbf{Edge Capacity:} 
For 5G networks, multiple-input, multiple-output (MIMO) technology is employed to deliver the capacity needs. This multi-antenna transmission technology boosts network capacity and data throughput significantly by increasing the number of transmission ports. The associated cost is that a higher number of processing resources are required for all the antenna signals. To transport all individual antenna signals to the baseband processing function, the transport capacity of the RAN network reduces. For this backhaul offloading scenario, edge networks play a major role in compensating for the over-consumption of network resources by extending the finite network resources, thus enhancing the delay and capacity performances of 5G networks \cite{9556285}. Therefore, artificial intelligence (AI) and associated techniques enable smart edge processing and open new opportunities for powerful computational processing,  massive data acquisition, and edge-caching capabilities requiring higher-order edge capacity. Hence, the vertical use case determines the required edge capacity. For instance, video streaming requires bulks of data caching, while for weather monitoring, small data updates over a period of time might be enough.  

\item   \textbf{Edge Intelligence:}
Traditional machine learning and deep learning techniques use higher volumes of training data and a large processing time, thus imposing storage and delay constraints on the centralized server. This is not a viable learning approach for B5G and 6G networks that promise massive connectivity with ultra-low latency. Edge intelligence accounts for powerful computational processing and massive data acquisition locally at edge networks \cite{zhang2022mobile}. It plays an important role in implementing smart and efficient resource scheduling strategies in a complex environment with heterogeneous resources and a massive number of devices. To this end, federated learning (FL) executes a training algorithm common to all edges to support decentralized learning further while keeping the data local. The result is a robust machine learning model able to address critical issues like data privacy, data security, data diversity, continual real-time learning, and hardware efficiency \cite{unal2021integration}.

\item\textbf{Security:} Although the EaaS approach of 5G-MEC improves the visibility over traditional networks and enables faster identification of potential security threats, the simultaneous operation of a massive number of IoT devices, cloud-based deployments, and third-party applications expands the attack surface. Edge computing provides enhanced security compared to cloud computing by allowing filtration of sensitive data at the source rather than sending it to a central data center. Therefore, fewer inter-device transfers of data using EaaS means better security. However, MEC is subject to vulnerabilities in virtualization technologies like virtual machine escape, virtual machine manipulation, DNS amplification, and virtual network function (VNF) location shift \cite{ranaweera2021mec}. Attacks on host-level orchestration components can collapse the MEC servers. Other possible security vulnerabilities and their mitigation strategies include detection of access to unauthorized files or applications, monitoring and blocking any information leakage, system anomaly detection, and prevention from over-utilization of host MEC resources. Using a privacy-preserving edge intelligence model like FL is currently being explored as an option to preserve data integrity by leveraging the distributed nature of edge-centric 5G verticals. FL ensures data privacy by executing the shared learning models locally and not sending the training data to central servers. 
\end{enumerate}
The key edge considerations are listed in Table \ref{tab:mec2}.

%% ====================================  Use Cases  ========

\section{B5G/6G Edge Networks - Use Cases}
This section provides an edge-centric overview of 5G applications and a projected view of B5G architectural requirements. 
%We present some major use cases, such as video analytics, intelligent transport systems (ITS), manufacturing, energy, and smart agriculture.

%% ==============  Video Analytics ================
\subsection{Video Analytics}
Video analytics aims to automatically recognize spatial and temporal events in videos, which is a data-intensive operation and gains particular advantage from the edge in terms of less data transfer time (less delay), local data processing (secure), and efficient resource utilization.

\subsubsection*{The Role of Edge Computing}
Massive and time-critical data processing are key requirements of video analytics. Low latency is primarily required for real-time content distribution and processing. Content caching of frequently used data at edge servers for on-request delivery to mobile devices is expected to reduce round-trip times (RTT) and provide high bandwidth in 5G wireless systems. However, the associated drawback is the limitation in processing capability at the edge compared to the data center's, therefore posing delays from software operation in a virtual environment. 
MEC has been integrated with 5G architecture to employ VNFs on the edge servers to overcome this delay.  Therefore, B5G networks can envision decentralized approaches like fog computing to reduce computing overhead further and reduce end-to-end delay.

\subsubsection*{Use Cases and Applications}
Some of the key B5G/6G verticals and use cases rely on video analytics for an operation like face recognition from traffic and security cameras, object tracking and motion detection for surveillance, AI learning of patterns in live stream video archives, etc. For example, the real-time experience of event audiences such as sports and concerts can be enhanced using video analytics at the edge. Augmented reality (AR) and virtual reality (VR) functions can be efficiently applied to videos captured from different camera angles and displayed through smartphones, tablets, screens, and AR/VR devices. Without edge, such data-intensive use cases cannot promise an enhanced user experience due to adversaries like delays and jitters. Some other vital application areas of video analytics are:

\begin{itemize}
\item \textbf{Smart City}. To enable smart cities, video analytics at the edge can play a key role. It can find applications in smart traffic monitoring, instant fire or smoke detection, waste disposal, and environmental monitoring.
\item \textbf{Surveillance}. Live video surveillance at the edge can instantly provide actionable results to avoid accidents and to enforce law and order.
\item \textbf{Autonomous Vehicles.} A self-driving car can maximize the benefit of edge video analytics using instantaneous data pre-processing for decisions and post-processing for continued training and feedback.
\end{itemize}

%% ==============  ITS  ================
\subsection{Intelligent Transport System (ITS)}
Intelligent transportation systems (ITS) have a great potential to make transportation systems safe, efficient, safe, and sustainable. IoT-driven intelligent transportation systems are defined by an amalgam of complex, data-intensive, dynamic, and uncertain operations like task management, resource allocation, and security that can be effectively managed using edge computing. 

\subsubsection*{The Role of B5G/6G and Edge Computing}
Future ITS applications are considered to be one of the most promising vertical sectors for mobile networks.  Two radio access technologies have been defined and are currently competing for market adoption: IEEE wireless access in vehicular environments (WAVE)/ dedicated short-range communications (DSRC) based on the IEEE 802.11p and 3GPP LTE cellular vehicle-to-everything (C-V2X) \cite{saeed2021wireless}. Since both these technologies rely on different channel access schemes, three major challenges arise when simultaneously accessing the spectrum in the 5.9 GHz band: coexistence, interoperability (for wide market acceptance), and backward compatibility.
The essence of B5G/6G is to satisfy a wide spectrum of heterogeneous end devices and vehicles requiring ultra-low latency and high bandwidth. This scale of computational power within proximity to the end-users can be made available at the edge of the network, using an edge cloud or edge data center (EDC). An edge cloud brings about a significant reduction in end-to-end latency in comparison with a centralized cloud. EDC networks face traffic burstiness, which is aggravated even further by the myriad of 5G access technologies. For example, even though the 5G mmWave offers massive peak-data-rate links, it makes the available capacity fluctuate. Therefore, the design of flexible EDC networks with promising ultra-low latency and high and scalable bandwidths is a promising research direction.

\begin{figure}
\centering
\includegraphics[width=0.95\columnwidth]{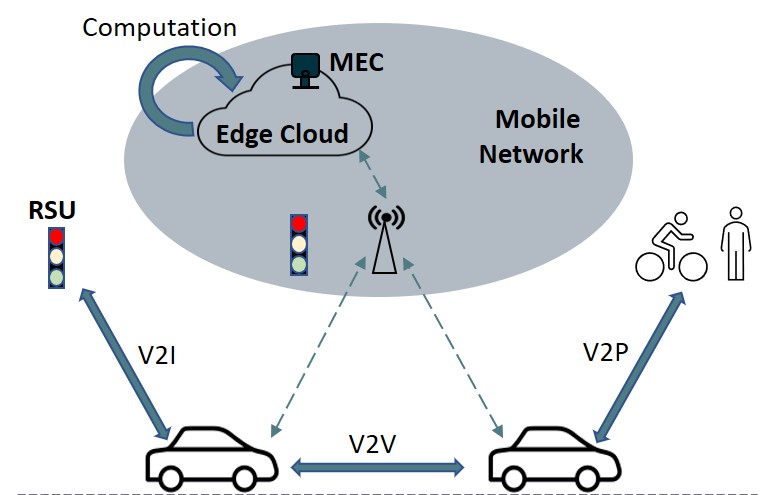}  
\caption{V2X system with multi-access edge computing.} 
\label{fig:its}
\vspace{-1.8em} 
\end{figure}

Fig. \ref{fig:its} shows a simple heterogeneous V2X system using MEC where ITS applications can communicate with road-side units (RSUs) and base stations. The MEC hosts can be co-located with these radiating elements (and/or with C-RAN aggregation points).

\subsubsection*{Applications and Use Cases}
ITS can be broadly categorized into three service areas \cite{9497103}:

\begin{itemize}
\item \textbf{Safety services.} To minimize accidents and mitigate risks for passengers and other users (pedestrians and vehicles).
\item \textbf{Non-safety services.} To improve traffic management to  maximize the efficiency of the existing road network and enhance experience by factors like traffic congestion minimization.
\item \textbf{Infotainment services.} To provide a range of services to the users of the vehicle, including Internet access, comfort services, and content sharing.
\end{itemize}

%%==================  Manufacturing ===============
\subsection{Smart Manufacturing}
B5G/6G networks will bring manufacturers and networks operators together to build smart factories capable of innovation, process automation and flexibility.  The current industrial networks have been designed for static manufacturing processes where even a single change in the manufacturing workflow requires long maintenance operations. With the fifth industrial revolution (Industry 5.0), the evolution of programmable network, increased security and dynamic traffic flows over heterogeneous industrial environments will redefine the manufacturing vertical of B5G connectivity. 
%This continuous evolution will undoubtedly unravel a drastic transformation in the industrial landscape, allowing innovative production models and promising new business opportunities. 

\begin{figure}
\centering
\includegraphics[width=0.9\columnwidth]{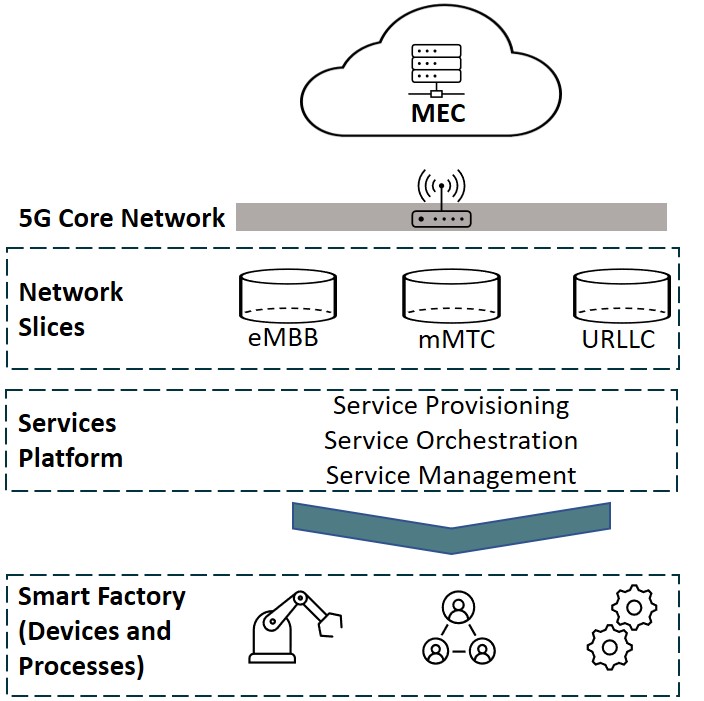}  
\caption{Process control for smart manufacturing.} \label{fig:manufacturing}
\vspace{-1.8em} 
\end{figure}

\subsubsection*{The Role of Edge Computing}
The success and sustenance of Industry 5.0 lies strongly in the adoption of emerging information and communications technology (ICT) technologies, such as loT, artificial intelligence, augmented reality for troubleshooting. By exploring access to real-time production information, cooperation and advanced control features, novel value added services are promised for industrial operators and customers. Edge computing and B5G/6G systems will play a key role in enabling Industry 5.0 by lending the network slicing paradigm for industrial use cases over a wide range of application areas.

Fig. \ref{fig:manufacturing} shows a layered architecture for process control in a smart factory or manufacturing facility. Each device and processes coordinates with the services platform for designing, creating and delivering the services. Platform layer is rendered dynamic by the network slicing layer which caters to 5G network resource management according to the service. Data management and storage is handled at the edge using MEC. To reduce the total average end-to-end transmission delay, efficient scheduling algorithms for backhaul links are required to route real-time traffic.

\subsubsection*{Applications and Use Cases}
Manufacturing vertical includes use cases like:

\begin{itemize}
\item  \textbf{Real time data analytics} finds important applications in safety procedures, product and process customization for output maximization and better resource utilization.
\item  \textbf{Augmented reality} can be used to determine operational inefficiencies to mitigate the cost of production downtime.
\item  \textbf{Object tracking} Real-time motion and robot tracking can overcome the tedious manual work and increase the efficiency of the manufacturing process.
\end{itemize}

%%==================  Energy  ===============
\subsection{Energy}
To keep up with United Nation's Sustainable Development Goals 2030, it is vital to ensure provision of clean energy for energy intensive sectors. 5G is an enabler for connected power distribution grids promising remote control and demand handling while minimizing energy wastage.

\subsubsection*{The Role of Edge Computing}
5G has a key role to address urgent issues like climate change and disaster monitoring, since it has a massive user base and can render energy usage efficient. To address the dynamic energy requirements of the 5G use cases, intelligent VPPs can ensure diversified resources, integrated energy consumption, production and trade. MEC has the potential to enable B5G with this capability. For instance, energy harvesting is a challenging task to supply it for low power devices at long ranges, since large aperture antennas are required. However, large antennas have a narrowing field of view, which limits their operation if they are widely dispersed from a 5G base station. The advantage of 5G edge centric cellular systems is that they are simultaneously truly ubiquitous, reliable, scalable, and cost-efficient. With smart integration of beam-forming techniques like \cite{eid2021}, B5G/6G itself can be an indispensable wireless power grid for IoT devices.

\subsubsection*{Applications and Use Cases}
Energy vertical includes use cases like:

\begin{itemize}
\item    \textbf{Virtual Power Plant (VPP)}
A virtual power plant (VPP) is a distributed cloud-based power plant that aggregates a high number of distributed energy resources (DERs), into an integrated and monitored network to enhance power generation. In a VPP, DERs are integrated into a single edge computing VPP network infrastructure. The objective of VPP is to dissipate power generated by individual units during peak load, thus relieving load on the smart grid. To achieve this, the VPP control center (centralized data center) collects data from edge devices like sensors, control units, and VPP applications. The edge computing gateway enables rapid response and real-time power dispatch. This allows predictive maintenance of the power grid by actively monitoring the operation of each power plant and energy storage.

\item  \textbf{Real time data analysis} To manage VPP services, optimal scheduling control of DERs is required in the VPP. In this realm, time-varying optimization of 5G can be employed for the real-time control of a VPP that could adjust the output of the DERs to maintain quality in the power network and achieve the objectives of both the customer and the utility company. This can reduce generation cost and power losses while increasing VPP service reliability.
\end{itemize}

\section{Experimental Analysis of Edge-based Services}
\begin{figure}[!h]
    \centering
    \includegraphics[width=0.99\columnwidth]{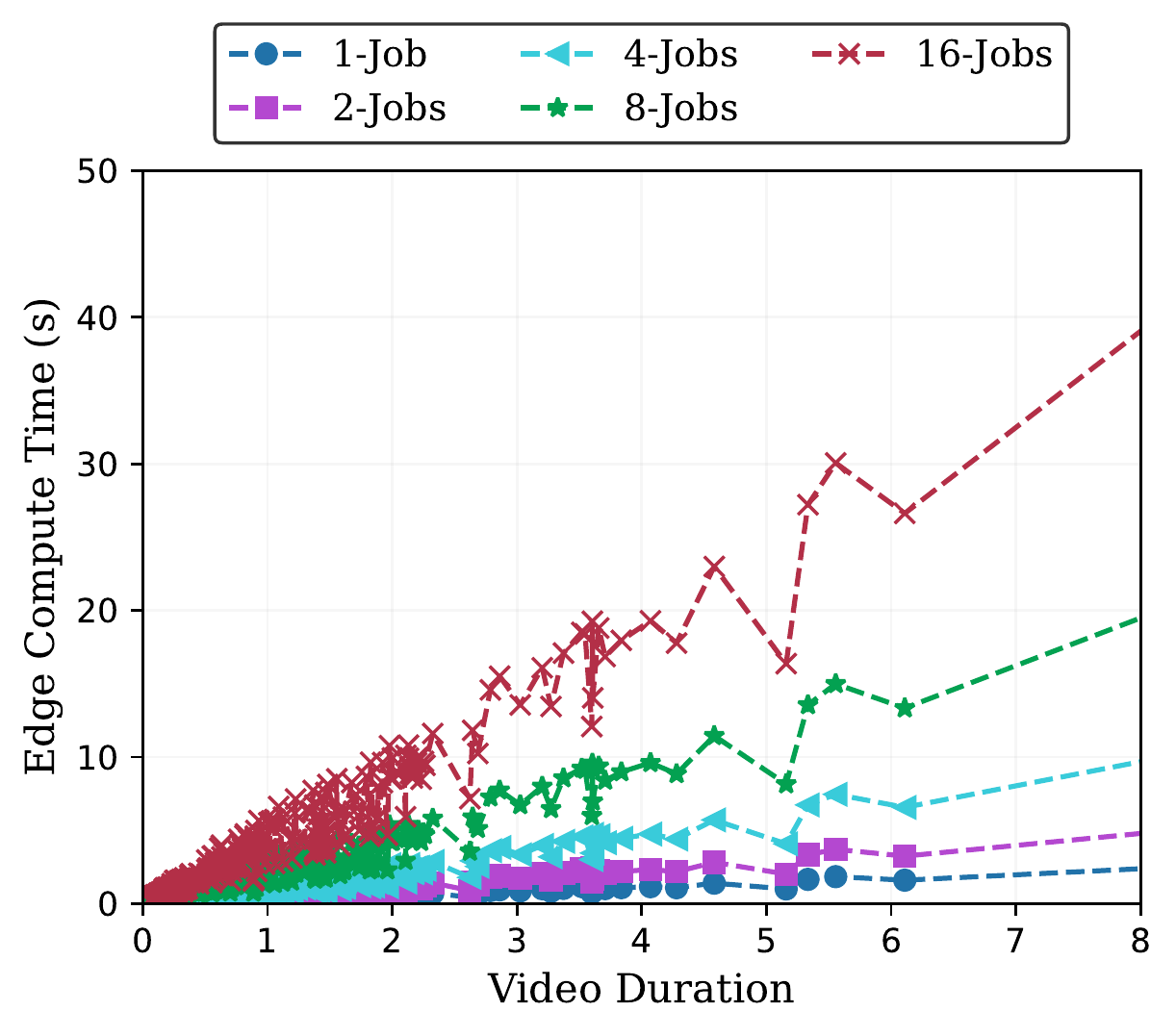}
    \caption{Edge-assisted video transcoding in various settings.}
    \label{fig:transcoding}
\end{figure}

This section presents some illustrative analysis of two edge-based services, including video transcoding and deep learning inference. Video transcoding is essential in real-time video streaming in content delivery and future VR/AR services. Edge servers can be used to enable fast and efficient transcoding services to enable these services. However, the limited processing capabilities of the edge servers can cause performance degradation when the number of simultaneous requests increases.
Fig. \ref{fig:transcoding} illustrates different scenarios where simultaneous transcoding jobs are executed at the edge server. The amazon web services (AWS)'s C5n provides a high-performance computing (HPC) platform for batch processing workloads, media transcoding, scientific modeling, machine learning inference, and other compute-intensive applications to transcode the videos. Videos of different sizes and different source bit rates are transcoded to evaluate the edge computing scenarios. Fig. \ref{fig:transcoding} depicts the edge computing time of transcoding videos of different duration, illustrating edge computing performance concerning the number of jobs. This intuitively motivates the use of efficient scheduling of jobs in edge computing platforms.

\begin{figure}[!h]
    \centering
    \includegraphics[width=0.99\columnwidth]{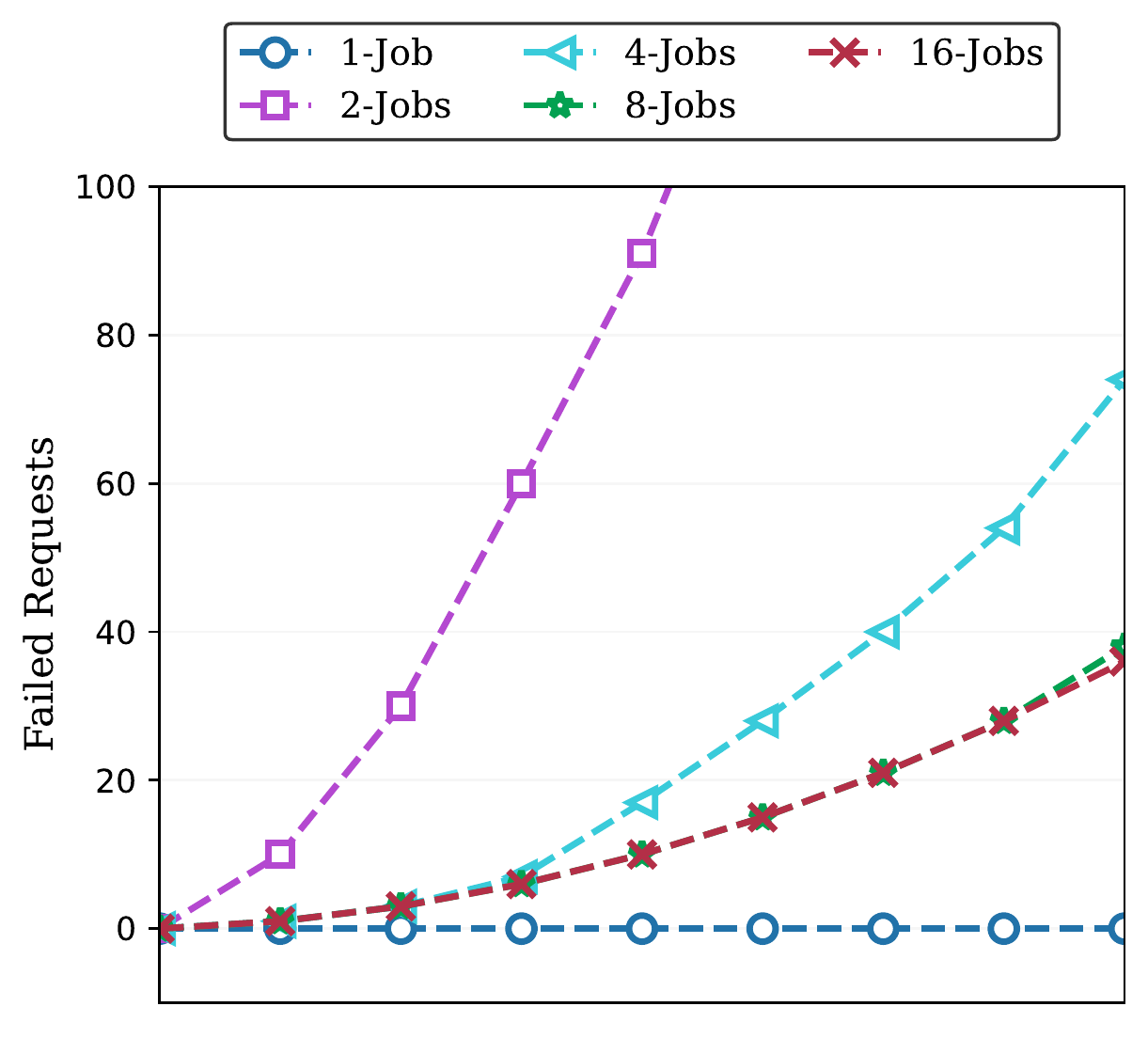}
    \caption{Failed processing requests.}
    \label{fig:transcoding2}
\end{figure}

We further analyze the computation power of an edge computing platform in  Fig. \ref{fig:transcoding2} which depicts the cumulative distribution (CDF) of the number of failed requests when using multiple simultaneous transcoding jobs. We consider the request as failed if it does not meet the time deadline (i.e., when the transcoding time exceeds the source video duration).

\begin{figure}[!h]
    \centering
    \includegraphics[width=0.99\columnwidth]{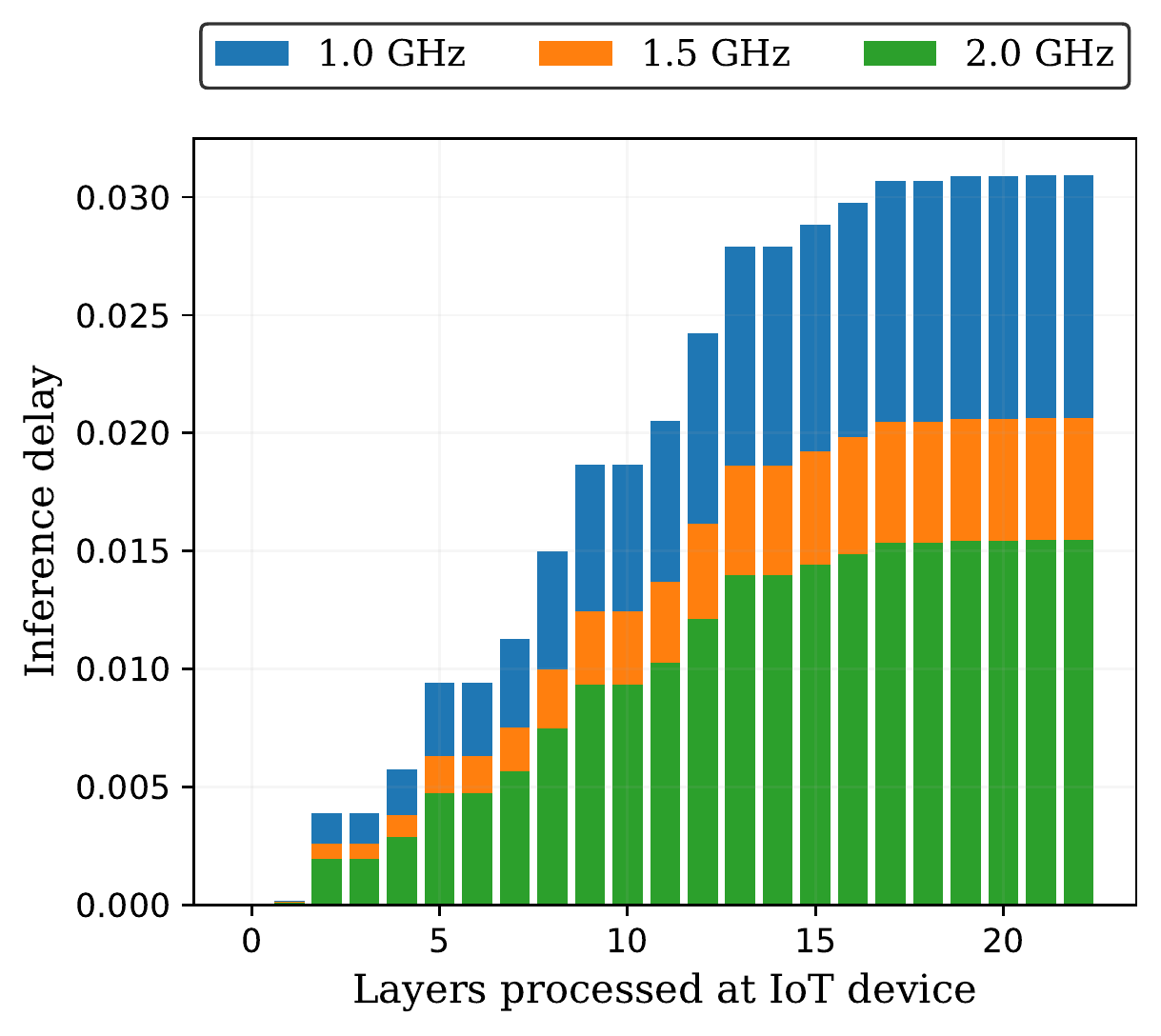}
    \caption{Inference delay using distributed edge computing.}
    \label{fig:inference_delay}
\end{figure}

While efficient scheduling of simultaneous jobs in the multi-edge environment can potentially solve the problem to a greater extent, the problem persists when only one or fewer edge servers are insufficient to execute the computations within the deadline. Collaboration among edge servers and end devices is a promising approach to address this challenge. Although such collaborations can use different setups, the standard practice is to partially execute the tasks using the local resources on the end device (mobile or IoT device) and partially offload to the edge server. Fig. \ref{fig:inference_delay} depicts our experimental analysis of performance inference delay while running the VGG16 deep learning model for image classification. The model is divided into two parts for each inference task depending upon the available local and remote computing resources and the available bandwidth of the wireless channel. The analysis is extended to use IoT devices of different CPU resources (i.e., 1.0 GHz, 1.5 GHz, and 2.0 GHz). The acquired results are representative of the VGG16 model, which indicates that for IoT devices of certain processing power, the inference delay increases when more layers are executed locally at the IoT device and potentially exceed a given threshold at a certain point (depending upon the application delay requirement). Hence, in such kind of collaborative environment, the processing tasks need to be optimally divided among IoT devices and the edge server.

%=========================================
%  Future Research Directions
%=========================================
\section{Future Research Directions}\label{futureWork}

Edge computing is an integral part of 5G IoT networks and envisioned 6G technologies that promise to provide more bandwidth with reduced latency to support novel cloud applications. In this paper, we presented key issues for edge computing in B5G/6G networks in terms of edge location, edge capacity, user density, local learning, and security. %Moreover, we also discussed major edge-centric B5G applications and services. 
In the following, we present some of the promising future research directions for edge computing in envisioned 6G networks.

\subsection{6G as Human Centric Networks} \label{subsec:hc_nwk}

6G networks shall be designed based on human-centric services that require tight-coupling of the QoE metrics to human senses rather than generic requirements. For example, 5G requires a latency requirement of 1 ms. In contrast, the 6G network provides latency based on the human reaction times, such as the 100 ms for auditory reaction, 10 ms for visual reaction, and 1 ms for perceptual response \cite{Yang2019}. Edge computing can play a vital role, thus providing guaranteed delay performance. The edge server has a more accurate local view of the network parameters and the users' instantaneous requirements; it becomes more convenient to adapt to the varying conditions and requirements when the application is running at the edge rather than the remote cloud server. Nevertheless, the research on investigating edge computing for 6G human-centric networks is still in infancy and needs the researcher's attention.

\subsection{Integrated Terrestrial and Aerial Networks} \label{subsec:integ_nwk}

B5G and 6G networks incorporate aerial and satellite networks as an essential part \cite{saeed2021point}. Aerial networks consisting of drones or unmanned aerial vehicles (UAVs) are considered as the first choice for applications such as search-and-rescue, surveillance, and wildlife conservation. These applications usually require real-time video analytics, which has several challenges in terms of communication (ultra-low latency), huge processing, and extra-large storage capacity. Although MEC in video applications brings significant improvement in latency reduction and computation offloading, live video analytics still poses challenges. Some key issues are (i) Computation migration from one MEC server to another when the drones fly away in overlapping coverage areas. (ii) Rapidly varying wireless link quality requires joint communication and computation and control (3C) services. (iii) Limited onboard energy of drones allows minimal local computation. (iv) D2D offloading of content and computations among drones on the fly to complement in areas of poor coverage or increasing demands than the current capacity of drones and/or MEC servers.  MEC-assisted drones and drone-drone cooperation in the video IoT (VIoT) services involve large-sized content sharing and intensive offloading, which intensify the challenges of normal drone cooperation.

\subsection{Self Sustained Networks (SSNs) using Edge Intelligence} \label{subsec:ssn_edge}

Another test case for 6G networks will be the use of artificial intelligence (AI) in the entire system architecture (i.e., from the network core to the edge). 6G networks should use AI-based services such as big data analytics for key decision-making in various real-time scenarios. Edge computing thus plays a significant role in 6G networks. The idea is that powerful edge servers can be deployed as an intermediate layer to provide localized and ultra-low latency processing for real-time applications. As cloud-based processing can suffer from long delays and transmission overhead, edge computing can complement cloud computing. The network services are designed such that fast content caching and processing are done at the edge to perform delay-sensitive critical tasks, whereas cloud servers are used to achieve deeper and more efficient processing at the expense of higher delays. However, edge resources are limited as compared to the cloud servers, and training AI models may be infeasible or inefficient. In this case, pervasive AI can be a promising solution that uses distributed network resources, including edge servers, cloud servers, and user devices, to intelligently distribute the AI computation (i.e., training and inference tasks) across various devices, to minimize the delay. Pervasive AI provides a further advantage in terms of privacy (due to no sharing of data) and scalability (using the huge distributed resources).

\subsection{Immersive Augmented and Mixed Reality Streaming}\label{subsec:ir}

6G networks would be able to meet the unprecedented requirements that 5G might not support. When we consider the huge volumes of data in future wireless services such as 8K videos, VR, AR, and even eXtended Reality (XR), the existing 5G support becomes insufficient. The limitation of 5G basically arises from the design and baseline KPIs that aim to reach average data rates of up to 100 Mbps (downlink) and 50 Mbps (uplink) [264]. In addition to the incapacity of 5G to deliver higher data rates for volumetric applications, 5G also does not meet the requirements of ultra-precision streaming tasks, such as remote surgeries and other XR applications. Therefore, investigating edge computing to support immersive augmented and mixed reality streaming is an exciting area of research.
\bibliographystyle{IEEEtran}
\bibliography{Manuscript_R2}
\vspace{-2.5 em}
%%%%% Authors Biographies
%\begin{IEEEbiography}[{\includegraphics[width=1in,height=1.25in]{{marium.jpg}}]
\begin{IEEEbiographynophoto}{Mariam Ishtiaq} received her Master's Degree in Electronics and Communication Engineering from Dongguk University, Seoul, South Korea. Her current research interests include underground sensor networks and applications of cloud computing and intelligence in 5G and B5G networks. Contact her at: mariam.ishtiaq@gmail.com
\end{IEEEbiographynophoto}
\vspace{-2.9 em}
%begin{IEEEbiography}[{\includegraphics[width=1in,height=1.25in]{{Nasir.jpg}}]{Naisr Saeed}
\begin{IEEEbiographynophoto}{Nasir Saeed}(S'14-M'16-SM'19) is currently an Associate Professor at Department of Electrical Engineering, Northern Border University, Arar, KSA.   His current areas of interest include localization, cognitive radio networks, underwater and underground communications, and aerial networks. Contact him at: mr.nasir.saeed@ieee.org
\end{IEEEbiographynophoto}
\vspace{-2.9 em}
% \begin{IEEEbiography}[{\includegraphics[width=1in,height=1.25in]{khan.jpg}}]{Muhammad Asif Khan}
\begin{IEEEbiographynophoto}{Muhammad Asif Khan}
is a Post-Doctoral Research Fellow at Qatar Mobility Innovation Center (a trade name of Qatar University QSTP-B). He received his Ph.D. degree in Electrical Engineering from Qatar University. His research focuses on mobile edge computing, distributed machine learning, and computer vision. He is a senior member of IEEE. Contact him at: asifk@ieee.org.
\end{IEEEbiographynophoto}

\vfill

\end{document}

%% file: tab_mec2.tex
\begin{table*}[h!]
\centering
\footnotesize
\caption{\label{tab:mec2}Edge considerations for vertical industries use-cases.}
 \begin{tabular}{ >{\columncolor{airforceblue!40}}m{3cm} >{\columncolor[gray]{0.9}}m{2.5cm} >{\columncolor{airforceblue!40}}m{2.5cm}  >{\columncolor[gray]{0.9}}m{2.5cm}  >{\columncolor{airforceblue!40}}m{2.5cm} >{\columncolor[gray]{0.9}}m{2.5cm} } 
 
 \textbf{Edge Considerations} & \textbf{Video Analytics} & \textbf{ITS} &\textbf{Manufacturing} & \textbf{Energy} & \textbf{Agriculture} \\ [0.5em] 
 \arrayrulecolor{white}\hline

%% 1 Architecture
Architecture &Single/Multi server & Multi server & Single server & Singe/Multi server & Multi server \\[1em] %\arrayrulecolor{gray!50}\hline

%% 2 Location
Edge location    &5G RAN    &RSU or RAN    &On-premise   & On-premise   &RAN\\[1em] %\arrayrulecolor{gray!50}\hline

%% 3 Storage Capacity
Storage capacity  &Very High   &Medium/High   &Low   &Low/Medium &Medium/High \\[1em] %\arrayrulecolor{gray!50}\hline

%% 3 Computational Capacity
Computing capacity &Very High   &High   &Low/Medium   &Medium  &Medium \\[1em] %\arrayrulecolor{gray!50}\hline

%% 4 Latency Requirements
Typical Latency Requirement & $<$ 1 s &10 ms - 1 s &$<$ 100 ms  &$<$ 10 ms & $<$ 1 s \\[1.5em] %\arrayrulecolor{gray!50}\hline

%% 6 Security
Security Issues &Data \& ML model poisoning &Data \& ML model poisoning  &Online adversarial attacks  &Online adversarial and DDoS attacks   &DDoS and Data poisoning  attacks \\[1em]
%\arrayrulecolor{gray!50}\hline

%% 5 Edge Intelligence
Edge Intelligence &Object detection, recommendation, summarization etc. &Object detection, tracking, traffic/vehicle prediction \& control etc. &Anomaly detection, process automation \& control etc. &Fault isolation, predictive maintenance etc.  &Anomalies detection, crop treatment recommendations etc. \\[1.5em]
%\arrayrulecolor{gray!50}\hline

%% Real-world deployments
Real-world deployments  &5G-ENSURE \cite{5Gensure}  &5G-ENSURE  &SONATA \cite{sonata} &COGNET \cite{CogNet}  &\rule{0pt}{4.2ex}5G-ENSURE

 \end{tabular}
\end{table*}
%------------------------------------------------------------------------------